\documentclass[11pt,preprint,aps,showpacs]{revtex4}
\usepackage{here}
\usepackage{graphicx}
\begin{document}

\title{ STRANGE PARTICLES AND NEUTRON STARS - \\
EXPERIMENTS AT GSI}
\author{P. Senger
\\
Gesellschaft f\"ur Schwerionenforschung, Planckstr. 1,\\
64291 Darmstadt, Germany}


\begin{abstract}
Experiments on strangeness production in nucleus-nucleus
collisions at SIS energies address fundamental aspects of modern
nuclear physics: the determination of the nuclear
equation-of-state at high baryon densities and the properties of
hadrons in dense nuclear matter. Experimental data and theoretical
results  will be reviewed. Future experiments at the FAIR
accelerator aim at the exploration of the QCD phase diagram at
highest baryon densities.
\end{abstract}
\maketitle

\section{Introduction}
\begin{figure} [hpt]
\vspace{0.cm}
\includegraphics[clip,width=13.cm]{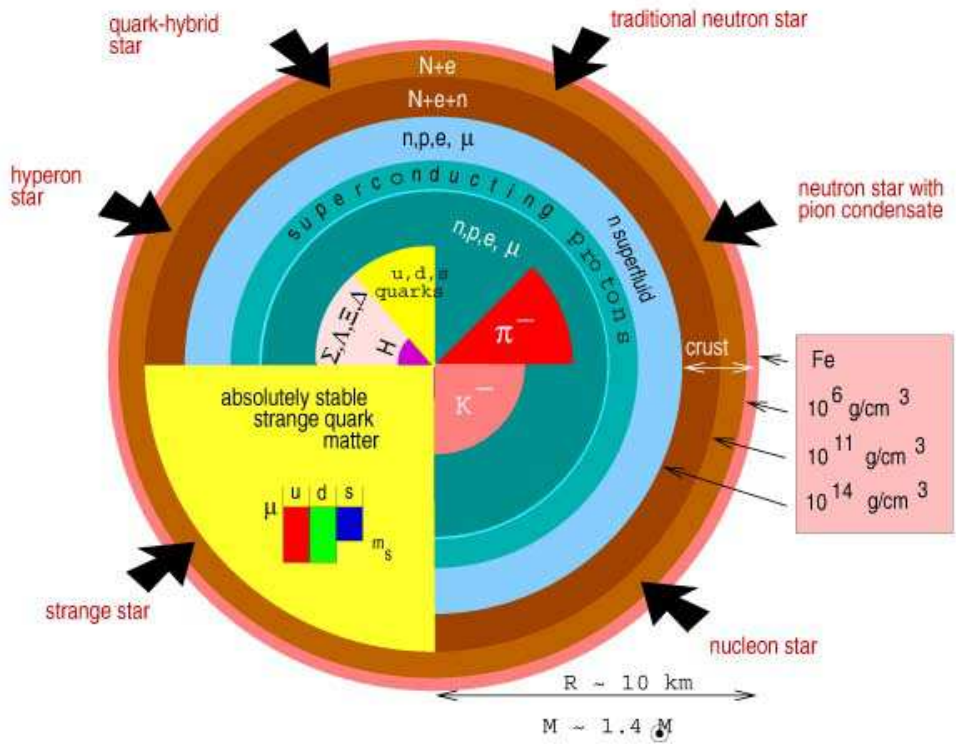}
\vspace{-0.cm} \caption{Structures of neutron stars  and novel
phases of subatomic matter as predicted by different models (taken
from \protect\cite{weber}) } \label{nstar}
\end{figure}

The goal of the nucleus-nucleus collision research program at the
present and the future GSI accelerators is to investigate the
properties of highly compressed nuclear matter. Such a form of
matter exists in various so far unexplored phases in the interior
of neutron stars and in the core of type II supernova explosions.
Figure \ref{nstar} illustrates structures of  neutron stars as
predicted by various models (compilation by F. Weber
\cite{weber}). As to date, none of these proposed novel phases of
subatomic matter can be ruled out by observation. Further
understanding of the structure on neutron stars requires more
experimental information on the nuclear equation-of-state at high
baryon densities, on  the in-medium properties of hadrons, in
particular of strange particles,  and on location of the
deconfinement phase transition at high baryon densities.

\subsection{The nuclear matter equation-of-state}
The nuclear matter equation of state plays an important role for
the dynamics of core collapse supernova and for the stability of
neutron stars. In type II supernova explosions, symmetric nuclear
matter is compressed to 2-3 times saturation density. These
conditions are realized in heavy-ion collisions at SIS18 beam
energies, although the temperatures reached in nuclear collisions
are higher than those in the core of a supernova. The isospin
asymmetric matter in the interior of neutron stars is compressed
to 5 - 10 times saturation density. At these high densities, the
nucleons are expected to melt into quarks and gluons. The
experimental study of this extreme state of matter in the
laboratory will be one of the major research programs at the
future  Facility for Antiproton and Ion Research (FAIR).

\begin{figure} [hpt]
\begin{minipage}[c]{0.6\textwidth}
\includegraphics[width=8.cm]{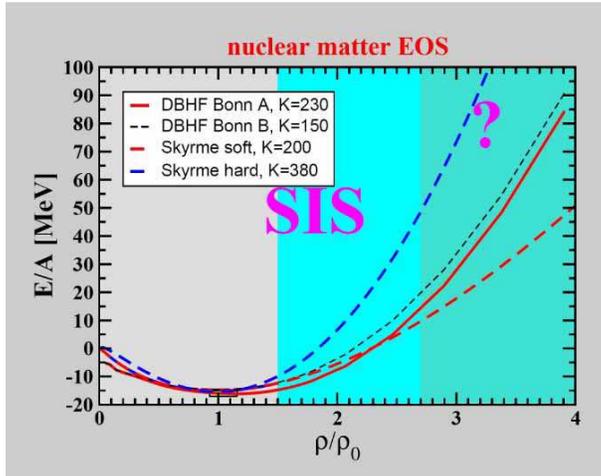}
\end{minipage}
\begin{minipage}[c]{0.35\textwidth}
\centering \caption{ Nuclear matter equations-of-state obtained
from relativistic Dirac-Brueckner Hartree-Fock calculations and
from a phenomenological model based on Skyrme forces. Both
approaches assume different values for the compressibility (see
\protect\cite{fuchs05}).}
 \label{eos}
\end{minipage}
\end{figure}

Figure \ref{eos} depicts different versions of the equation of
state as predicted by different calculations (see \cite{fuchs05}).
The figure illustrates that it is not sufficient to determine the
nuclear compressibility (which is determined by the curvature of
E($\rho$) at saturation density), but rather one has to study the
response of nuclear matter at different densities, which means one
has to perform nucleus-nucleus collisions at different beam
energies.

The study of the equation-of-state (EOS) of symmetric nuclear
matter is one of the most challenging goals of the nuclear
collision experiments at GSI.  The experimental observables
related to the EOS are the collective flow of nucleons and the
production of strange particles. Microscopic transport
calculations indicate that the yield of kaons created in
collisions between heavy nuclei at subthreshold beam energies
(E$_{beam}$ = 1.58 GeV for NN$\to$$K^+\Lambda$N) is sensitive to
the EOS of nuclear matter at high baryon densities
\cite{aich_ko,li_ko}. This sensitivity is due to the production
mechanism of K$^+$ mesons. At subthreshold  beam energies, the
production of kaons requires multiple nucleon-nucleon collisions
or secondary collisions  such as $\pi$N$\to$$K^+\Lambda$. These
processes are expected to occur predominantly at high baryon
densities, and the densities reached in the fireball depend on the
nuclear equation-of-state \cite{fuchs97}. According to transport
calculations, in central Au+Au collisions the bulk of K$^+$ mesons
is produced at nuclear matter densities larger than twice
saturation density.

Moreover, $K^+$ mesons are well suited to probe the properties of
the dense nuclear medium because of their long mean free path. The
propagation of  $K^+$ mesons in nuclear matter is characterized by
the absence of  absorption (as they contain an antistrange quark)
and hence kaons emerge as messengers from  the dense phase of the
collision. In contrast, the pions created in the high density
phase of the collision are likely to be reabsorbed and most of
them will leave  the reaction zone in the late phase.

The influence of the medium on the $K^+$ yield is amplified by the
steep excitation function of kaon production near threshold
energies. Early transport calculations find that the $K^+$ yield
from Au+Au collisions at subthreshold energies will be  enhanced
by a factor of about 2 if a soft rather than a hard
equation-of-state is assumed \cite{aich_ko,li_ko}. Recent
calculations take into account the modification of the  kaon
properties in the dense nuclear medium \cite{ko_li,cass_brat}(see
next chapter). When  assuming a repulsive $K^+$N potential as
proposed by various theoretical models (see \cite{schaffner} and
references therein) the energy needed to create a $K^+$ meson in
the  nuclear medium is increased  and hence the $K^+$ yield will
be reduced. Therefore, the yield of $K^+$ mesons  produced in
heavy ion collisions is affected by both the nuclear
compressibility and the in-medium kaon potential.

\begin{figure}[H]
\begin{minipage}[c]{0.55\textwidth}
\includegraphics[width=8.cm]{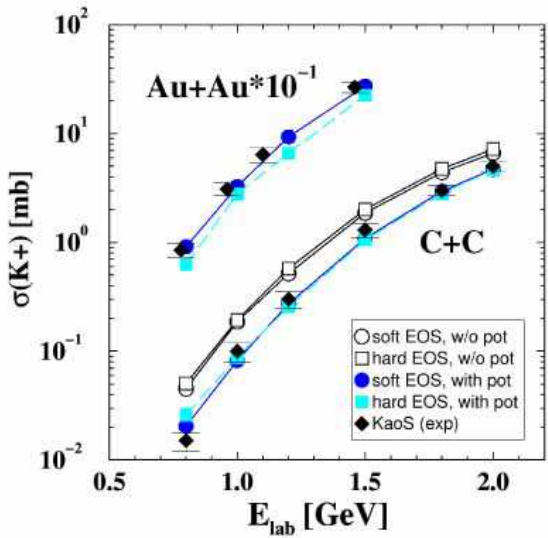}
\end{minipage}
\begin{minipage}[c]{0.4\textwidth}
\centering\caption{Production cross sections of $K^+$ mesons  for
Au+Au and C+C collisions as a function of the projectile energy
per nucleon. The data (full diamonds) are compared to results of
transport calculations assuming a soft (circles) and a hard
(squares) nuclear equation-of-state with and without $K^+N$
in-medium potentials. Taken from   \protect\cite{fuchs05}. }
\label{eosexci}
\end{minipage}
\end{figure}
The KaoS collaboration proposed to disentangle these  two
competing effects by studying $K^+$ production in a very light
($^{12}$C+$^{12}$C) and a heavy collision system
($^{197}$Au+$^{197}$Au) at different beam energies near threshold
\cite{sturm01}. The reaction volume is more than 15 times larger
in Au+Au than in C+C collisions and hence the average baryonic
density - achieved by the pile-up of nucleons - is significantly
higher \cite{cass_brat}. Moreover, the maximum baryonic density
reached in Au+Au collisions depends on the nuclear compressibility
\cite{li_ko,aichelin} whereas in the small C+C system this
dependence is very weak \cite{fuchs01}.

\begin{figure} [H]
\begin{minipage}[c]{0.55\textwidth}
\centering\includegraphics[width=8.cm]{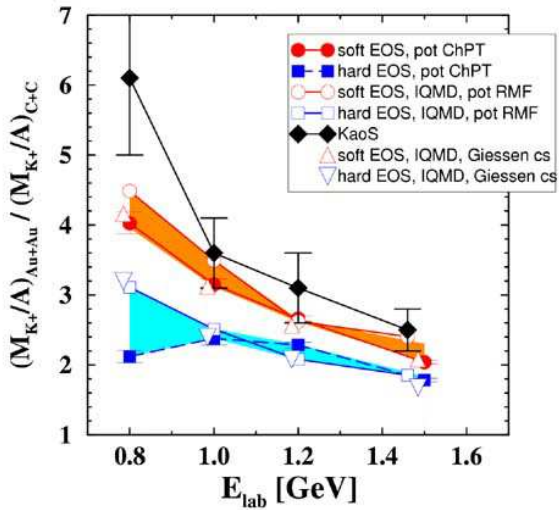}
\end{minipage}
\begin{minipage}[c]{0.4\textwidth}
\centering\caption{K$^+$ ratio measured in inclusive Au+Au and C+C
collisions as function of beam energy \protect\cite{sturm01}. The
data are compared to various QMD calculations assuming  nuclear
compressibilities of $\kappa$ = 200 MeV  and 380 MeV . Figure
taken from \protect\cite{fuchs05}. } \label{karatio}
\end{minipage}
\end{figure}

Recent QMD transport calculations which  take into account a
repulsive kaon-nucleon potential reproduce the energy dependence
of the kaon ratio  if a compression modulus of $\kappa$ = 200 MeV
for nuclear matter is assumed \cite{fuchs01}. These calculations
use momentum-dependent Skyrme forces to determine the
compressional energy per nucleon (i.e. the energy stored in
compression) as function of nuclear density (see figure
\ref{eos}). The result of this calculation is presented in figure
\ref{eosexci} in comparison with the data \cite{sturm01}.

In order to reduce systematic uncertainties both in experiment
(normalization, efficiencies, acceptances etc.) and theory
(elementary cross sections etc.) the K$^+$ multiplicities are
plotted as ratios (K$^+$/A)$_{Au+Au}$ / K$^+$/A)$_{C+C}$ in figure
\ref{karatio}. In this representation also the in-medium effects
cancel to a large extent. The calculations are performed with a
compression modulus of $\kappa$ = 380 MeV (a ''hard''
equation-of-state) and with $\kappa$ = 200 MeV (a ''soft''
equation-of-state) \cite{fuchs05,fuchs01}. The data clearly favor
a soft equation of state (see figure \ref{karatio}).

\subsection{In-medium properties of kaons and antikaons}

According to various calculations, the properties of kaons and
antikaons are modified in dense baryonic matter (see e.g.
\cite{cass_brat,schaffner,brown91}). In mean-field calculations,
this effect is caused by a repulsive $K^+$N potential and an
attractive $K^-$N potential. As a consequence, the total energy of
a kaon at rest in nuclear matter increases and the antikaon energy
decreases with increasing density. It has been speculated that an
attractive $K^-$N potential will lead to Bose condensation of
$K^-$ mesons in the core of neutron stars above baryon densities
of about 3 times saturation density \cite{li_lee_br}.

\begin{figure} [htp]
\centering\includegraphics[width=11.0cm]{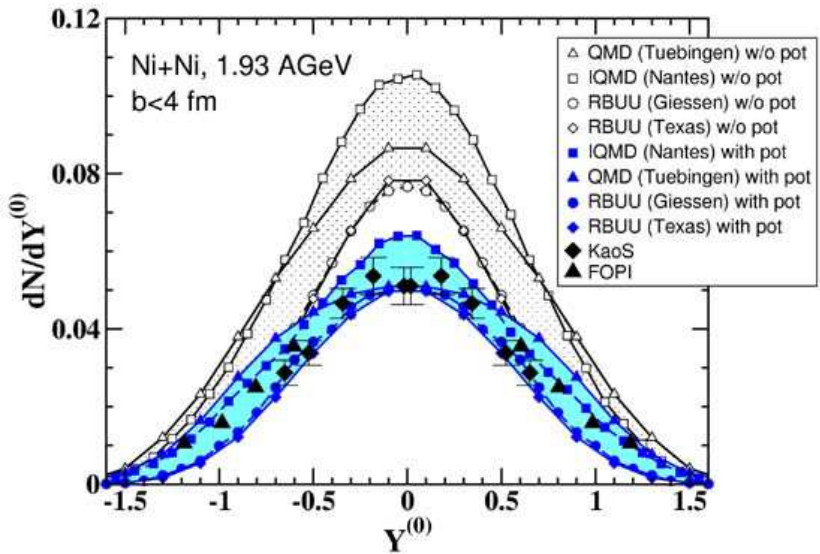}
\caption{Multiplicity density distributions of $K^+$ mesons for
near-central (b $<$ 4.4 fm) Ni+Ni collisions at 1.93 AGeV. Full
black diamonds: KaoS data \protect\cite{menzel}, full black
triangles: FOPI data \protect\cite{best,wisnie}.  The data are
compared to various transport calculations. Full symbols: with
in-medium effects. Open symbols: without in-medium effects. Taken
from \protect\cite{fuchs05} } \label{dndy}
\end{figure}

A comparison of experimental data to results of various transport
model calculations is presented in figure~\ref{dndy}. The figure
shows the $K^+$ multiplicity densities dN/dy for near-central
Ni+Ni collisions at 1.93 AGeV as function of the c.m. rapidity.
Figure~\ref{dndy} combines data measured by the KaoS Collaboration
\cite{menzel} and by the FOPI Collaboration \cite{best,wisnie}.
The transport calculations clearly disagree with the data if
in-medium effects are neglected (open symbols). However, when
taking into account a repulsive K$^+$N potential, the $K^+$ yield
is reduced, and the calculations agree with the data sufficiently
well (full symbols).

Another observable consequence of in-medium $KN$ potentials is
their influence on the propagation of kaons and antikaons in
heavy-ion collisions. The measured  azimuthal emission patterns of
$K^+$ mesons  contradict the expectations based on a long mean
free path in nuclear matter. The particular feature of sideward
flow \cite{crochet} and the pronounced out-of-plane emission
around midrapidity \cite{shin} indicate that $K^+$ mesons are
repelled from the regions of increased baryonic density as
expected for a repulsive $K^+N$ potential \cite{li_ko_br,wang99}.

\begin{figure} [htp]
\centering\includegraphics[width=11.0cm]{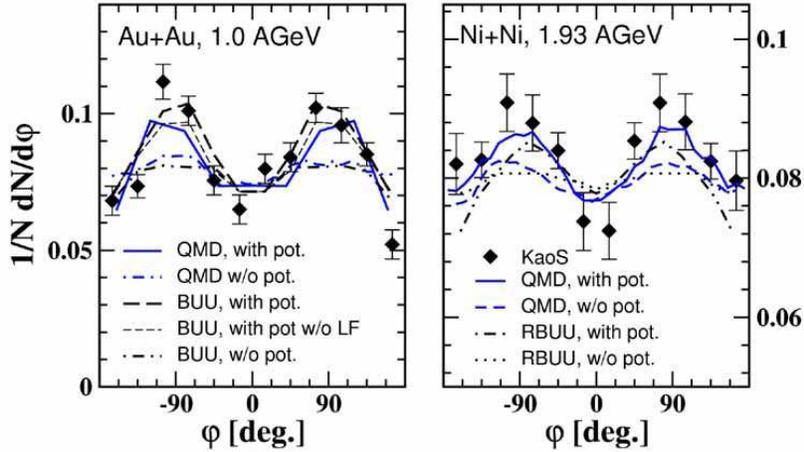}
\caption{$K^+$ azimuthal distributions for semi-central Au+Au at 1
AGeV (left panel) and for Ni+Ni collisions at 1.93 AGeV (right
panel). The data (full symbols) are compared to various transport
calculations with and without in-medium potentials (see insert).
Taken from \protect\cite{fuchs05}, see also
\protect\cite{larionov}. } \label{auni_squeeze_qmd}
\end{figure}

Figure \ref{auni_squeeze_qmd} depicts the azimuthal angle
distributions of $K^+$  mesons measured in Au+Au collisions at 1.0
AGeV (left panel) and in Ni+Ni collisions at a beam energy of 1.93
AGeV (right panel). The $K^+$ emission patterns clearly are peaked
at $\phi$=$\pm$90$^0$ which is perpendicular to the reaction
plane. Such a behavior is known from pions which are shadowed by
the spectator fragments. In the case of $K^+$ mesons, however, the
anisotropy can be explained by transport calculations only if a
repulsive in-medium $K^+N$ potential is assumed. A flat
distribution is expected when neglecting the in-medium potential
\cite{fuchs05,larionov}.

\begin{figure}[htp]
\begin{minipage}[c]{0.5\textwidth}
\centering\includegraphics[width=5.cm]{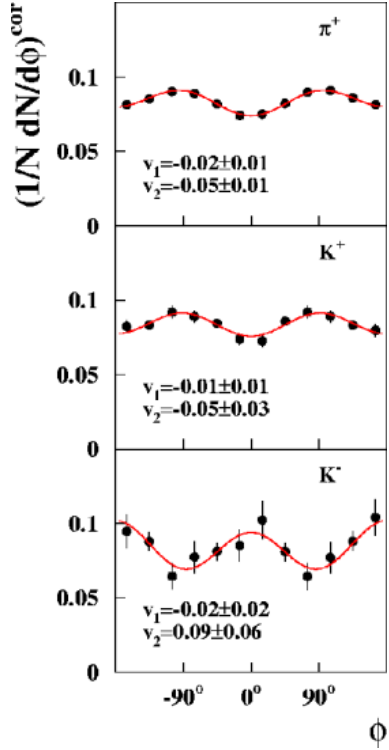}
\end{minipage}
\begin{minipage}[c]{0.4\textwidth}
\caption{$\pi^+$, $K^+$ and $K^-$ azimuthal distribution for
semi-central Ni+Ni collisions at 1.93 AGeV \protect\cite{uhlig}.
The data are corrected for the resolution of the reaction plane
and refer to impact parameters of 3.8 fm$ < b <$ 6.5 fm,
rapidities of 0.3 $< y/y_{\rm beam} <$ 0.7 and momenta of  0.2
GeV/c $< p_t <$ 0.8 GeV/c. The lines represent Fourier expansions
fitted to the data. The resulting coefficients are indicated. }
\label{nisqueeze}
\end{minipage}
\end{figure}

The key observable for the $K^-N$ in-medium potential is the
azimuthal emission pattern of K$^-$ mesons. The existence of an
attractive K$^-$N potential will strongly reduce the absorption of
$K^-$ mesons, and consequently  the K$^-$ mesons will be emitted
almost isotropically in semicentral Au+Au collisions
\cite{wang99}. This observation would be in contrast to the
behavior of pions and K$^+$ mesons, and would provide strong
experimental evidence for in-medium modifications of antikaons.
Figure \ref{nisqueeze} depicts the first data on the $K^-$
azimuthal emission pattern measured in heavy-ion collisions (lower
panel) which differs significantly from the corresponding $K^+$
and pion pattern (center and upper panel) \cite{uhlig}. First IQMD
calculations with and without in-medium potentials for $K^+$ and
$K^-$ mesons are able to reproduce the data \cite{hart01}. Further
clarification will come from high statistics data which have been
measured in Au+Au collisions at 1.5 AGeV, and which are presently
being analyzed.

In the mean field calculations as discussed above the K$^-$ mesons
are treated as quasi-particles which are on the mass shell.
Microscopic coupled-channel calculations based on a chiral
Lagrangian, however,  predict a dynamical broadening of the $K^-$
meson spectral function in dense nuclear matter
\cite{waas,lutz03}. First off-shell transport calculations using
$K^-$ meson spectral functions have been performed
\cite{cass_tolo}. The ultimate goal of the calculations is to
relate the in-medium spectral function of $K^-$ mesons to the
anticipated chiral symmetry restoration at high baryon density.
New experimental information on the in-medium modification of
vector mesons is expected from the dilepton experiments with HADES
at GSI. Highest baryon densities will be produced and explored
with the Compressed Baryonic Matter (CBM) experiment at the future
FAIR accelerator center in Darmstadt.

\section{Towards highest baryon densities}
The future international Facility for Antiproton and Ion Research
(FAIR) in Darmstadt will provide unique research opportunities in
the fields of nuclear, hadron, atomic and plasma physics
\cite{cdr}. The accelerators will deliver primary beams (protons
up to 90 GeV, Uranium up to 35 AGeV, nuclei with Z/A = 0.5 up to
45 AGeV) and secondary beams (rare isotopes and antiprotons) with
high intensity and quality. The aim of the nucleus-nucleus
collision research program is to explore the QCD phase diagram at
high net baryon densities and moderate temperatures. This approach
is complementary to the studies of matter at high temperatures and
low net baryon densities performed at RHIC and LHC. At high baryon
densities, new phases of strongly interacting matter are expected
\cite{weber,wilczek,stephanov}. This exciting new field of high
baryon density QCD needs input from experimental data, which can
only be provided by new and dedicated  nucleus-nucleus collision
experiments.

\begin{figure} [H]
\begin{minipage}[c]{0.45\textwidth}
\center\includegraphics[width=6.cm]{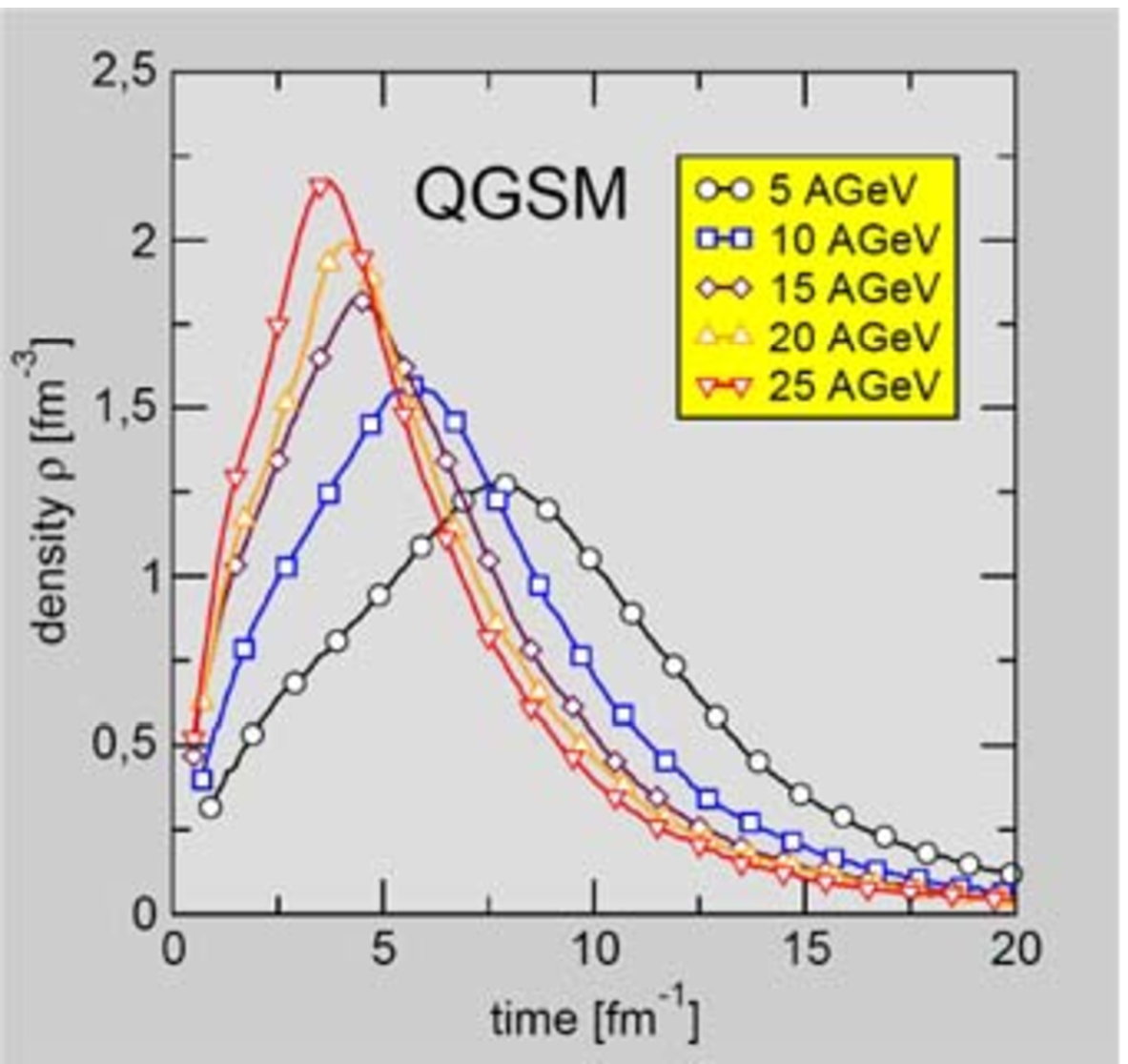}
\end{minipage}
\begin{minipage}[c]{0.45\textwidth}
\center\includegraphics[width=6.cm]{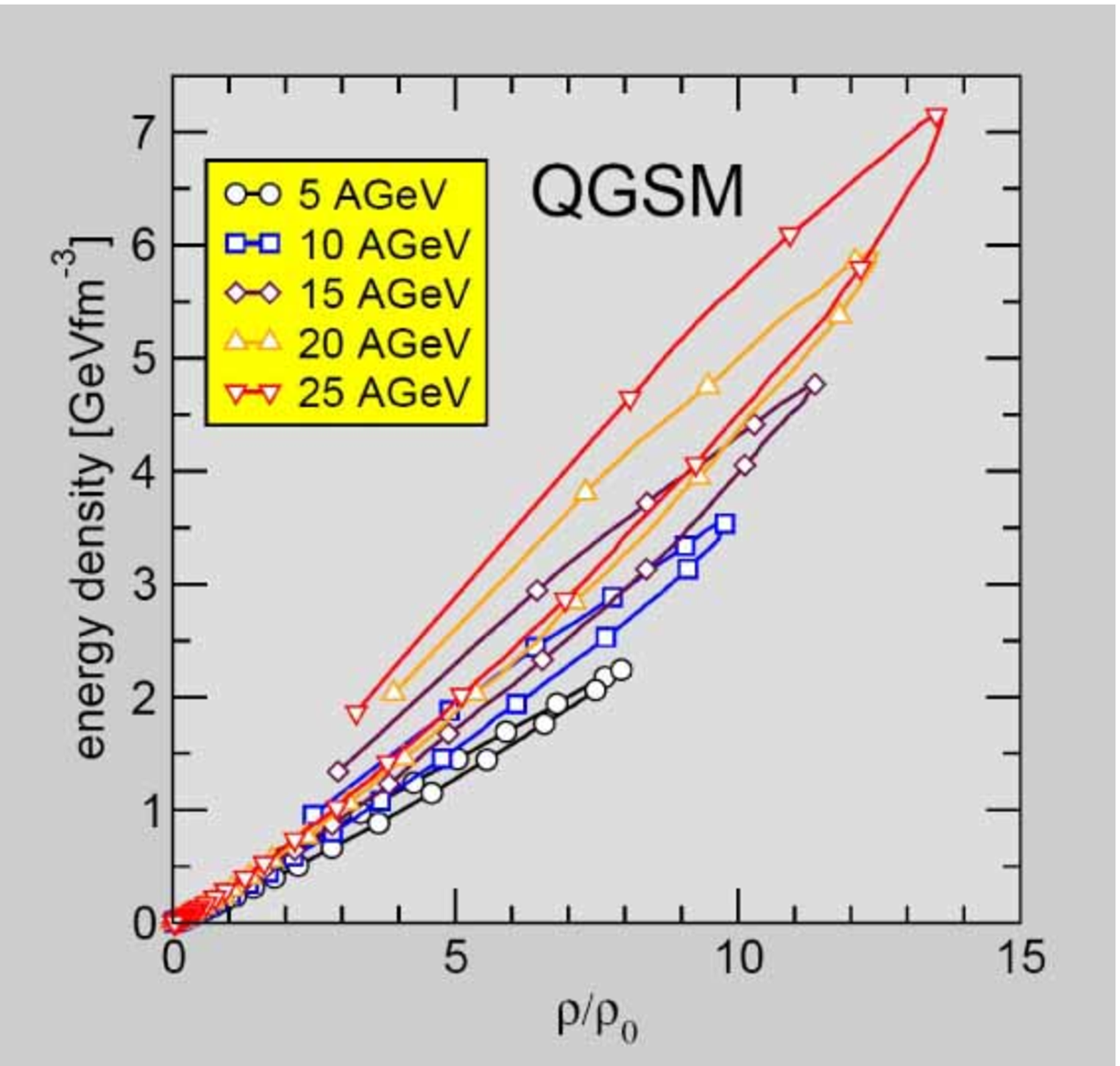}
\end{minipage}
\caption{ Left panel: nuclear density in the inner volume of
central Au+Au collisions as function of time calculated with the
Quark Gluon String Model QGSM  for beam energies between 5 and 25
AGeV \protect\cite{burau}. Right panel: Corresponding energy
density as function of the baryon density (in units of saturation
density).} \label{transport}
\end{figure}

Very high baryon densities - which are comparable to those in the
core of neutron stars - are predicted to be reached in heavy-ion
collisions already at moderate beam energies. This is illustrated
in figure \ref{transport} (left panel) which depicts the density
in the inner volume of  central Au+Au collisions as function of
time calculated with a transport code (quark gluon string model
QGSM, see \cite{burau}). Already at a beam energy of 5 AGeV the
baryon density exceeds a value of 1 fm$^{-3}$, i.e. more than 6
times saturation density. The corresponding energy density is
plotted in the right panel of figure \ref{transport} as function
of baryon density (in units of saturation density). At a beam
energy of 5 AGeV the energy density is predicted to reach a value
of 2 GeV fm$^{-3}$ which is - according to lattice QCD
calculations - already beyond the deconfinement phase transition.

\begin{figure} [htp]
\centering\includegraphics[width=9.cm]{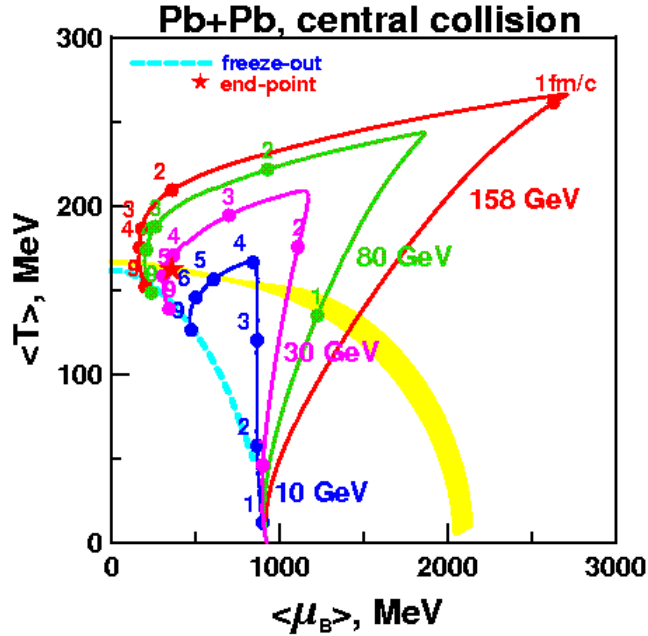}
\caption{Trajectories of heavy ion collisions in the QCD phase
diagram calculated by a 3-fluid hydrodynamics  model
\protect\cite{ivanov}. Dashed line: chemical freeze-out curve.
Star: critical endpoint.  } \label{hydro}
\end{figure}

Trajectories of nucleus-nucleus collisions in the T - $\mu_B$
plane have been calculated with a 3-fluid hydrodynamics model
\cite{ivanov}. Figure \ref{hydro} depicts  the trajectories
corresponding to different beam energies. The phase diagram also
contains the critical endpoint (star) which has been predicted
recently by lattice QCD calculations \cite{fodor,ejiri}. The
lattice calculations find a first order phase transition above
$\mu_B \approx$ 400 MeV, and a smooth cross over from hadronic to
partonic matter below this value. The hydrodynamics model
calculation indicates that this critical endpoint might be found
in the vicinity of the freeze-out point of a central Pb+Pb
collision at a beam energy of about 30 AGeV.

The CERN-NA49 collaboration found a pronounced peak in the
excitation function of the K$^+/\pi^+$ ratio in central Pb+Pb
collisions at a beam energy of 30 AGeV \cite{NA49}. This structure
cannot be explained by any theoretical model.  The Lambda/pion
ratio exhibits also a maximum at the same beam energy. In this
case, the shape of the excitation function can be reproduced by a
statistical model analysis, which indicates that the transition
from baryon dominated matter to meson dominated matter occurs at a
beam energy of about 30 AGeV.  It turns out that the maximum
net-baryon density at freeze-out is reached at a beam energy of 30
AGeV \cite{cleymans06}.

\subsection{Experimental observables}

The major challenge is to find diagnostic probes which are
connected to the onset of chiral symmetry restoration, to the
deconfinement phase transition, and to the equation of state of
hadronic and partonic matter.

The in-medium spectral functions of short-lived vector mesons -
which are expected to be sensitive to chiral symmetry restoration
-  can be studied in the dense nuclear medium via their decay into
lepton pairs \cite{rapp}. Since the leptons are very little
affected by the passage through the high-density matter, they
provide, as a penetrating probe, almost undistorted information on
the conditions in the interior of the collision zone. Another
observable sensitive to in-medium effects is open charm, e.g.
D-mesons. The effective masses of D-mesons - a bound state of a
heavy charm quark and a light quark - are expected to be modified
in dense matter similarly to those of kaons. Such a change would
be reflected in the relative abundance of charmonium (c$\bar{c}$)
and D-mesons.

The onset  of a first order phase transition is expected to cause
a discontinuity in the excitation function of particular
observables. Such a non-monotonic behavior has been observed
around 30 AGeV in the kaon-to-pion ratio and in the inverse slope
parameter of kaons \cite{NA49}. A beam energy scan looking at a
variety of observables  is needed to clarify the experimental
situation. This includes the measurement of the phase-space
distributions of strange particles, in particular multi-strange
baryons (anti-baryons), and particles containing charm quarks. For
example, a discontinuity in the excitation function of the
J/$\psi$-to-$\psi$' ratio would indicate sequential charmonium
dissociation due to color screening in the deconfined phase.
Moreover, event-by-event fluctuations are expected to appear when
crossing a first order phase transition, and particularly in the
vicinity of the critical endpoint. The identification of a
critical point would provide direct evidence for the existence and
the character of a deconfinement phase transition in strongly
interacting matter.

The formation of a mixed phase indicating the onset of
deconfinement leads to a softening of the equation of state at a
given beam energy \cite{hung}. The location of the so called
softest point (i.e. where the sound velocity exhibits a minimum)
may be discovered by measuring carefully the excitation function
of the collective flow of particles.

Charm production plays a particular role at FAIR energies, because
charmonium, D-mesons and charmed hyperons are created at beam
energies close to the kinematical threshold. Therefore, these
particles are sensitive probes of the early, high-density stage of
the collision (more than 10 times saturation density !).
Collective effects contributing to charm production may be visible
for the first time. Charm exchange processes may become important,
revealing basic properties of charm propagation in a dense
baryonic medium. The situation is analogue to strangeness
production at SIS18 energies, where 2-3 times saturation density
is probed in Au+Au collisions. In order to perform high statistics
measurements, the low cross sections for charm production at
threshold beam energies has to be compensated by high beam
intensities.

\subsection{The CBM detector} 
\begin{figure}[H]
\center\includegraphics[width=16.cm]{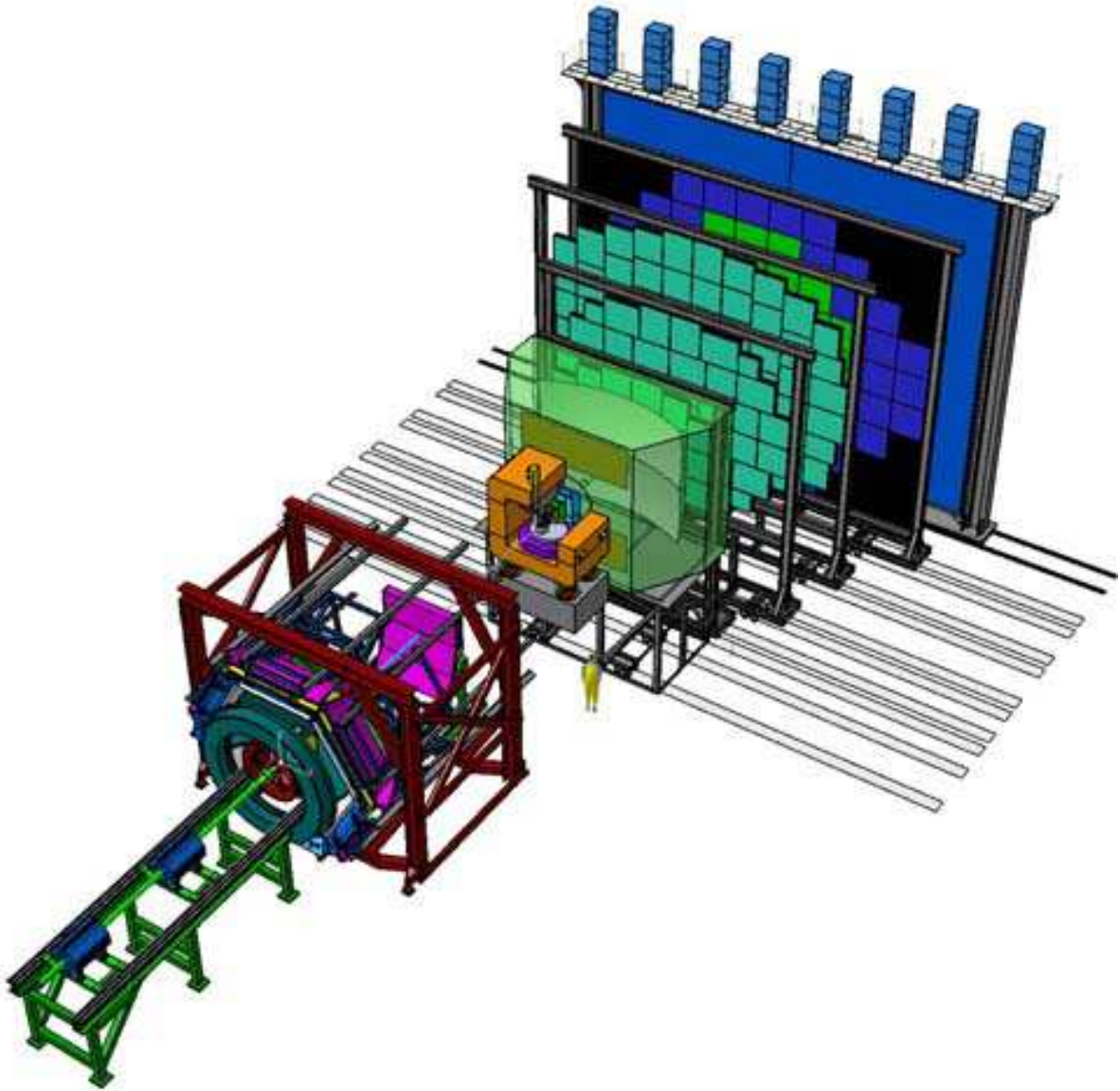}
\caption{Sketch of the planned Compressed Baryonic Matter (CBM)
experiment together with the HADES detector.  } \label{cbm}
\end{figure}

The experimental task is to identify
both hadrons and leptons and to detect rare probes in a heavy ion
environment. The apparatus has to measure multiplicities and
phase-space distributions of hyperons, light vector mesons,
charmonium and open charm (including the identification of
protons, pions and kaons) with a large acceptance.  The challenge
is to filter out those rare probes in Au+Au (or U+U) collisions at
reaction rates of up to 10$^7$ events per second. The charged
particle multiplicity is about 1000 per central event. Therefore,
the experiment has to fulfill the following requirements: fast and
radiation hard detectors, large acceptance, lepton and hadron
identification, high-resolution secondary vertex determination and
a high speed trigger and data acquisition system. The layout of
the CBM experimental setup is sketched in figure~\ref{cbm}.

The CBM setup consists of a the following subsystems:
\begin{itemize}
\item a large acceptance superconducting dipole magnet \item a
radiation-hard Silicon Tracking Station comprizing  pixel/strip
detectors \item a Silicon pixel microvertex detector with high
position resolution and low material budget \item  a Ring Imaging
Cherenkov detector (RICH) for soft electron identification \item
Transition Radiation Detectors (TRD) for identification of
electrons with momenta above 2 GeV/c \item  a muon detection
system consisting of several layers of absorbers and tracking
chambers. This system is an alternative to RICH and TRDs, which
then would be used as tracking stations for hadron identification.
\item  Resistive Plate Chambers (RPC) for time of flight
measurement (hadron identification) \item an Electromagnetic
calorimeter (ECAL) for identification of photons \item a high
speed online event selection and data acquisition system
\end{itemize}

The CBM detector is designed for a comprehensive research program
using proton beams (with energies of 10 - 90 GeV)  and nuclear
beams (10 - 45 AGeV) impinging on various targets. The
measurements at beam energies below 10 AGeV will be performed with
the HADES detector. The measurements, in particular those of rare
diagnostic probes, require a dedicated accelerator with high beam
intensities, large duty cycle, excellent beam quality, and with an
operational availability of several month per year. Details of the
CBM research program and of the setup can be found in  the
Technical Status report which has been submitted in January 2005
\cite{CBM}. The CBM Collaboration actually consists of about 350
persons from 41 institutions and 15 countries.
\section*{Acknowledgements}
I would like to thank my colleagues of the KaoS collaboration who
have measured most of the data on strangeness production at SIS
presented in this paper: I.~B\"ottcher, M.~D\c{e}bowski,
F.~Dohrmann, A.~F\"orster, E.~Grosse, P.~Koczo\'n, B.~Kohlmeyer,
F.~Laue, M.~Menzel, L.~Naumann, H.~Oeschler, M. Ploskon,
W.~Scheinast, E.~Schwab, Y.~Shin, H.~Str\"obele, C.~Sturm,
G.~Sur\'owka, F.~Uhlig, A.~Wagner, W.~Walu\'s.

The CBM project is supported by EU under RII3-CT-2004-506078
HADRONPHYSICS and by INTAS Ref. Nr. 03-51-6645.


\begin{thebibliography}{60}

\bibitem{weber} F. Weber, J. Phys. G:, Nucl. Part. Phys. {\bf 27} (2001)
465
\bibitem{fuchs05} C. Fuchs, Prog.Part.Nucl.Phys. 56 (2006) 1
\bibitem{aich_ko} J. Aichelin, C. M. Ko, Phys. Rev. Lett. {\bf 55} (1985) 2661
\bibitem{li_ko} G. Q. Li, C. M. Ko, Phys. Lett. {\bf B 349} (1995) 405
\bibitem{fuchs97}  C. Fuchs et al., Phys. Rev. {\bf C 56} (1997) R606
\bibitem{ko_li} C.M. Ko and G.Q. Li, J. Phys. {\bf G 22} (1996) 1673
\bibitem{cass_brat}  W. Cassing and E. Bratkovskaya, Phys. Rep. {\bf 308} (1999) 65
\bibitem{schaffner} J. Schaffner-Bielich, J. Bondorf, I. Mishustin,
Nucl. Phys. {\bf A 625} (1997) 325
\bibitem{sturm01} C. Sturm et al., Phys. Rev. Lett. {\bf 86} (2001) 39
\bibitem{aichelin}  J. Aichelin, Phys. Rep. {\bf 202} (1991) 233
\bibitem{fuchs01} C. Fuchs et al., Phys. Rev. Lett. {\bf 86} (2001) 1974
\bibitem{brown91} G.E. Brown et al., Phys. Rev. {\bf C 43} (1991) 1881
\bibitem{li_lee_br} G.Q. Li, C.H. Lee and G.E. Brown, Phys. Rev. Lett. {\bf 79} (1997) 5214
\bibitem{menzel} M. Menzel et al., Phys. Lett. {\bf B 495} (2000) 26
\bibitem{best} D. Best et al., Nucl. Phys. {\bf A 625} (1997) 307
\bibitem{wisnie} K. Wisniewski et al., Eur. Phys. J. {\bf A 9} (2000) 515
\bibitem{crochet} P. Crochet et al., Phys. Lett. {\bf B 486} (2000) 6
\bibitem{shin} Y. Shin et al., Phys. Rev. Lett. {\bf 81} (1998) 1576
\bibitem{li_ko_br} G.Q. Li, C.M. Ko,  G.E. Brown, Phys. Lett. {\bf B 381} (1996)
17
\bibitem{wang99} Z.S. Wang et al., Eur. Phys. J. {\bf A 5} (1999) 275
\bibitem{larionov} A. Larionov and U. Mosel, Phys. Rev. C 72 (2005) 014901
\bibitem{uhlig} F. Uhlig et al., Phys. Rev. Lett. {\bf 95} (2005) 012301
\bibitem{hart01} C.~Hartnack et al., Eur.~Phys.~J.~ {\bf A 1} (1998) 151
\bibitem{waas} T. Waas, N. Kaiser and W. Weise, Phys. Lett. {\bf B 379} (1996)
34
\bibitem{lutz03} M. F. M. Lutz and  C. Korpa, Nucl. Phys. {\bf
A 700}(2002) 209
\bibitem{cass_tolo}  W. Cassing et al., Nucl.Phys. {\bf A 727} (2003) 59
\bibitem{cdr} An International Accelerator Facility for Beams of Ions and Antiprotons,
Conceptional Design Report 2001, \begin{verbatim}
http://www.gsi.de/GSI-Future/cdr/ \end{verbatim}
\bibitem{wilczek} F. Wilczek, Physics Today {\bf 53} (2000) 22
\bibitem{stephanov} M. Stephanov, K. Rajagopal, E. Shuryak, Phy.
Rev. {\bf D 60} (1999) 114028
\bibitem{burau} G. Burau et al., Phys.Rev. C71 (2005) 054905
\bibitem{ivanov}  Y. Ivanov, V. Russkikh, V.Toneev, Phys.Rev. C73 (2006) 044904
\bibitem{fodor} Z. Fodor and S.D. Katz, JHEP {\bf 0404} (2004) 050
\bibitem{ejiri} S. Ejiri et al., hep-lat/0312006, Prog.Theor.Phys.Suppl. 153 (2004) 118
\bibitem{NA49} C. Blume et al., J. Phys. {\bf G 31} (2005) 685
\bibitem{cleymans06} J. Cleymans et al., Phys. Rev. {\bf C 73}
(2006) 034905
\bibitem{rapp} R. Rapp and J. Wambach, Nucl. Phys. {\bf A 661} (1999) 33c
\bibitem{hung} C. M. Hung and E. V. Shuryak, Phys. Rev. Lett. {\bf
75} (1995) 4003
\bibitem{CBM} CBM Technical Status Report 2005 \begin{verbatim}
http://www-linux.gsi.de/~hoehne/report/cbmtsr_public.pdf
\end{verbatim}
\end{thebibliography}
\end{document}